\documentclass[prb,floats,superscriptadress,twocolumn,showpacs,preprintnumbers,amsmath,amssymb]{revtex4}

\usepackage[dvips]{graphicx}
\usepackage{graphicx}
 
\usepackage{dcolumn}
 
\usepackage{bm}

\usepackage{rotating}

\newlength{\La} 

\newlength{\Lb} 

\newlength{\Lc} 

\newcolumntype{d}{D{.}{.}{-1}}

\renewcommand{\arraystretch}{1.25}

\newcommand{\grad}{\ensuremath{^\circ}}

\newcommand{\cgo}{CuGeO$_3$}

\newcommand{\sppu}{spin-Peierls transition}

\newcommand{\tzp}{T$_{2}^+$}

\newcommand{\tsp}{T$_{SP}$}

\newlength{\figwidth}
 
\divide\figwidth by 2
 
\begin{document}

\title{Lattice dynamics of CuGeO$_3$ : inelastic neutron scattering and model calculations}
\author{M.~ Braden$^{a,b,c,*}$,W. Reichardt$^a$, , B. Hennion$^b$,
G. Dhalenne$^d$, A. ~Revcolevschi$^{d}$}
\address{
$^a$ Forschungszentrum Karlsruhe, IFP, Postfach 3640, D-76021
Karlsruhe, Germany\\
$^b$Laboratoire L\'eon Brillouin,
C.E.A./C.N.R.S., F-91191-Gif-sur-Yvette CEDEX,
France\\
$^c$ II. Phys. Inst., Univ. zu K\"oln, Z\"ulpicher Str. 77, D-50937 K\"oln, Germany\\
$^d$Laboratoire de Physico-Chimie de l'Etat Solide,
Universit\'e Paris Sud, 
91405 Orsay Cedex, France 
}
\date{\today}

\begin{abstract}
The lattice dynamics in CuGeO$_3$ has been analyzed by the combination 
of inelastic neutron scattering studies 
and lattice dynamical model calculations.
We report an almost complete set of dispersion curves along
the three orthorhombic directions and along [101]. 
The dispersion of branches associated with the 
modes directly involved in the spin-Peierls transition
allows to explain the particular propagation vector of the
structural distortion in the dimerized phase. 

\end{abstract}


\maketitle

\section{Introduction}

The spin-Peierls transition in CuGeO$_3$ has attracted considerable interest
due to the relatively simple crystal structure of this material
allowing experimental studies so far not possible in the organic
spin-Peierls compounds
\cite{hase93a}. In particular neutron scattering and diffraction
techniques could be used for the study of this material
\cite{pouget94,regnault96a,hirota94}.

The spin-Peierls transition is based on magneto-elastic coupling :
the dimerization of the magnetic chain results from the structural distortion.
Since the gain in magnetic energy is linear, and since the loss in
elastic energy is quadratic in the distortion, in a simple one-dimensional
picture the combined structural and magnetic transition must occur.
In real systems inter-chain coupling may, however, favor
three-dimensional antiferromagnetic ordering.

In CuGeO$_3$, the microscopic coupling has been clarified by experimental --
magnetic and  structural --  and theoretical studies in great detail.
The spin-$1/2$ chains in CuGeO$_3$ are formed by the CuO$_2$-chains characterized by
edge sharing CuO$_4$; the oxygen in these chains is labeled
O2 \cite{mueller-buschbaum89}.
These chains are directly connected in
the $b$-direction by GeO$_4$-tetrahedra.
In contrast, the coupling along $a$ through the apical oxygen, O1,
is only weak due to the long CuO1-distance.
A detailed description of the crystal structure and its temperature dependence
is given in references \cite{braden96,braden98a}.
The variation of the magnetic interaction parameter, 
$J$, 
in the spin-Peierls phase results mainly from the modulation
of the Cu-O-Cu bond-angle, $\eta$, 
which in contrast to the high T$_c$ superconductor parent compounds
is close to 90\grad ~  \cite{braden96,geertsma96a}.
In an isolated CuO$_2$-chain with 90\grad ~ bond angle, there should be
no antiferromagnetic exchange, since two exchange-paths cancel each other.
In CuGeO$_3$ the exact cancellation of these rather large values is destroyed
by the deviation of the bond-angle from 90\grad ~ and by the hybridization
with the Ge-atoms acting as side-groups.
This renders the magnetic interaction quite sensitive to small
structural changes.
Already in the non-dimerized phase the magneto-elastic coupling
causes anomalous temperature dependences via the equilibrium of
magnetic and structural energy\cite{winkelmann95,braden98a}.
In the spin-Peierls phase the dimerization is achieved by the modulation
of both the Cu-O-Cu bond angle $\eta$, and  
the angle between the CuO$_2$-ribbons
and the Ge-O1-bond, $\delta$ \cite{braden96}.
The quantitative relation between the bond angles and $J$ has been
calculated by several techniques :
on the base of band structure calculations \cite{braden96,geertsma96a,feldkemper}, 
from thermodynamic considerations \cite{buechner98iv}
and within a RPA-theory of the spin-Peierls transition\cite{werner98}.
A comparative discussion of the distinct techniques is given in ref. 
\cite{werner98}.

It is obvious that phonons play an important role in any
spin-Peierls transition.
Therefore, the lattice dynamics of CuGeO$_3$ has been studied by several techniques.
Due to its transparency CuGeO$_3$ presents
favorable conditions for optical methods; almost all zone-center
frequencies have been determined by the combination of
Raman and Infrared studies \cite{popovic95b}.
However, the spin-Peierls transition is characterized by a breaking
of translational symmetry, therefore, the phonons directly involved
in the transition must have propagation vectors away from $\Gamma$.
In CuGeO$_3$ the lattice is doubled along $a$ and $c$, the propagation vector
is, hence, (0.5 0 0.5). 
Several groups have challenged the
question whether the involved phonon modes soften close to the
transition or not 
by inelastic neutron scattering
\cite{lorenzo94,hirota95b,nishi95c}.
These studies tried to solve the problem by measuring only a few
branches, which does not allow to characterize the lattice dynamics 
of this complex system. 
CuGeO$_3$ has 10 atoms in its primitive cell, the associated
lattice dynamics with its 30 phonon branches is close to the maximum
complexity one may treat today.
The identification of the involved phonon modes could be 
achieved
only by the complete study reported here.

We have already published results concerning the phonon modes
directly related to the spin-Peierls transition \cite{braden98b}. 
There are four modes
with the symmetry of the transition from Pbmm to Bbcm.
Two aspects of these modes were unexpected. First, the distortion of the
dimerized phase does not correspond to just one of these modes, but at least
two polarization patterns have to be combined.
Second, the two strongly involved modes do not soften close to the
transition in contradiction with the Cross-Fisher prediction \cite{cross79a}.
These results have stimulated new theoretical studies :
Gros and Werner have extended the Cross-Fisher theory in order
to explain the missing of phonon softening \cite{gros98pp}
and Uhrig \cite{uhrig98a} has stressed
the importance of the high phonon frequencies compared to
the magnetic energy scale, i.e. a non-adiabatic condition.
In general it has been shown that coupling of the spin system
with a branch of phonons is important for a quantitative understanding
of the spin-Peierls transition \cite{wellein, bursill}.

In this paper we want to present the general lattice dynamical study
furnishing the basis for the identification of the relevant phonon modes.
Only the combination of neutron scattering with lattice dynamical
calculations permits to treat a phonon problem of this complexity.
CuGeO$_3$ is one of the few complex systems where the lattice dynamics may
be considered as being understood in detail. In this sense it is a rather
promising material for validating ab-initio procedures to
calculate phonon frequencies, so far restricted to less complex
systems. In addition, the lattice dynamics of CuGeO$_3$ presents several
aspects not related to the spin-Peierls scenario but interesting
in themselves. Furthermore, the phonon dispersion of CuGeO$_3$ should
have some exemplary character for the wide class of silicates and
germanates \cite{liebau85} 
and also for other cuprate chain systems of current
interest.

In section II we describe the neutron scattering
studies and present the lattice dynamics of CuGeO$_3$.
The phenomenological model used to accompany the measurements and
to interpret the data is introduced in section III.
The general discussion of the phonon dispersion in CuGeO$_3$ based on
the experimental studies and the model is given in section IV.
In section V we give some further information on the modes directly
involved in the transition which is not included in our previous paper.
Finally, in section VI, we briefly discuss the possibility of 
distinct structural order parameters.

\section{Inelastic neutron scattering studies}

The main part of the studies of the phonon dispersion
has been performed on the triple axis spectrometer 1Ta,
operated by the Forschungszentrum Karlsruhe at the Orph\'ee reactor
in Saclay.
A few additional studies have been made on the spectrometer
2T also installed at the Orph\'ee reactor \cite{llbblau}.

Since it has been necessary to study the phonon dispersion in
CuGeO$_3$ in its entity, the amount of beam time spent on this
problem was considerable.
25 days on 1T and 7 days on 2T were used in order to determine
an almost complete set of phonon branches along the three orthorhombic
directions and along  [101].
For the study of the temperature dependence of modes relevant
to the spin-Peierls transition, another 8 days were used on 1Ta.

For a first experiment, two crystals of 550 mm$^3$ volume each
have been coaligned in the  [100]/[001]-orientation using a double goniometer.
Due to the long $b$-axis this orientation allows attaining
Q-values out of the scattering plane by tilting the sample with the
goniometer.
Along the [100]-direction almost all modes could be determined in
this orientation.
One of the two crystals shows a larger mosaic spread around $a$, which was
tolerable in the [100]/[001]-orientation since
it becomes hidden by the large vertical divergence.
For orientations with $a$ not parallel to the scattering plane
it has been advantageous to use only one crystal.
Further experiments were performed in the [010]/[001]-, [100]/[010]-
and [101]/[010]-orientations.

For all measurements a pyrolytic graphite (PG) analyzer was used;
in the low frequency range up to $\sim$10THz a PG-(002) monochromator
and, at high frequencies, a Cu-(111) monochromator.
For a few studies requiring an exceptional resolution
a Cu-(220) monochromator was mounted.
Due to double focusing arrangement of monochromators and analyzers
the spectrometer 1T yields a considerable gain in intensity 
\cite{pintschovius}. 
The focusing arrangement is incompatible with the use of
collimators but yields a better resolution in comparison to
a configuration with flat crystals and open collimations.
Roughly, the focusing configuration corresponds to a conventional
spectrometer with 37'-collimators throughout.
All scans were performed with the final energy fixed
in the neutron energy loss mode and, apart a few
scans aiming at acoustic phonons, the final energy was
fixed to the values of 3.555 and 7.37THz, where
higher order contaminations may be suppressed by the PG filter.

The phonon dispersion was studied mainly at room temperature;
just for the highest frequency modes it has been favorable
to cool the sample to 10K in order to reduce the background.

In figure 1 we show typical phonon scans which illustrate the
different effort to be made to analyze phonon frequencies at
low and high energy. The intensity, $I$, of a one phonon process
in neutron energy loss mode is given by \cite{squires}:

$$
I \propto \cdot {1\over \omega} 
\cdot (n(\omega)+ 1) 
\{\sum_{d}{b_d \over \sqrt{m_d}} \cdot 
e^{(-W_d+i{\bf Q}\cdot {\bf r_d})} 
\cdot({\bf Q}\cdot {\bf e_d})\}^2  ~~~~~~(1),$$

($\omega$ denotes the phonon frequency, ${\bf q}$ the wave-vector,
${\bf Q}={\bf g} +{\bf q}$ the scattering vector (${\bf g}$ the reciprocal lattice vector)
$e^{-W_d}$ the Debye-Waller-factor,
and the sum extends over
the atoms in the primitive cell with mass $m_d$ scattering
length $b_d$, position ${\bf r_d}$ and polarization vector ${\bf e_d}$.
In general,  the intensity is determined by the Bose-factor, $n(\omega)$,
and the $1\over \omega$-term,
which strongly reduce the effectiveness to observe
high energy modes.
In analogy to the elastic structure factor, the sum in (1) is called
dynamic structure factor; it describes the interference of the
interaction of the distinct atoms weighted by the scalar product
of their displacements with the scattering vector ${\bf Q}$.
Only phonons with some polarization parallel to ${\bf Q}$ can be observed.
The right side of equation (1) may be calculated with a lattice dynamical
model in order to predict favorable conditions for the measurement
and in order to identify single phonon modes.

\begin{figure}[t,p]
\resizebox{0.7\figwidth}{!}{  
\includegraphics*{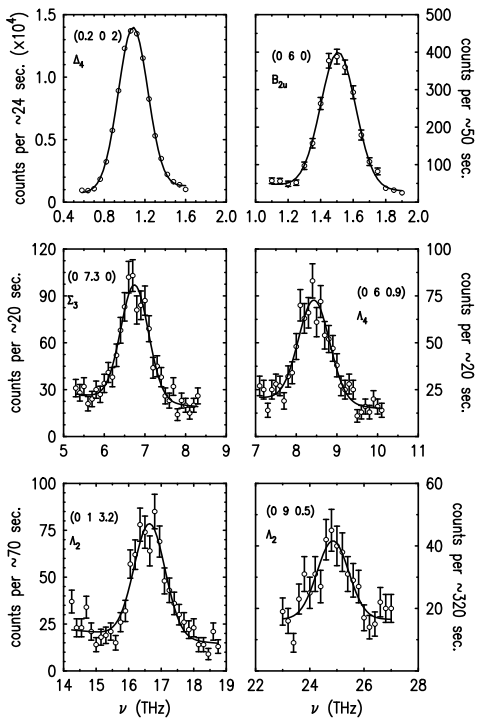}}
\caption{Exemplary scans aiming at determining phonon frequencies
at the  $\Gamma$-point and in the Brillouin-zone with fits by Gaussian
profiles and sloping background.}
\end{figure}

Figure 1 shows a scan across a transverse acoustic phonon at
${\bf g}$=(0 0 2) (upper left); the mode is polarized along the $c$-direction.
This mode yields a high peak intensity in accordance with the
strong (0 0 2) Bragg-reflection intensity.
In contrast, the determination of the highest frequencies
requires an extreme effort in beam-time; the figure 1 (lower right)
presents results for a strong dynamic structure factor;
modes with unfavorable structure factors may not be analyzed
in practice in this frequency region.
In the medium energy range the analysis of the phonon
spectra is dominated by the question whether the modes
may be isolated for a certain ${\bf Q}$-value.

The phonon frequencies were obtained by fitting Gaussian distributions
to the measured profiles. This procedure may be used
in principle only when the curvature of the dispersion surface
can be neglected in the range of the
four-dimensional resolution ellipsoid.
The whole set of dispersion curves along the orthorhombic directions,
given in figure 2,
is separated according to the irreducible representations.
Only a few high energy branches were not studied in our experiments; however,
their zone center frequencies are fixed by the optical techniques and the
dispersion is most likely flat.

\section{Lattice dynamical model calculations for CuGeO$_3$}

In order to analyze a complex problem like the phonon dispersion
it is essential to take profit of symmetry considerations.
The 10 atoms in the primitive cell correspond to 30
zone-center frequencies characterized by their polarization vectors.
The symmetry of the CuGeO$_3$ crystal structure allows a separation
corresponding to irreducible representations \cite{rousseau}
each of them being characterized by a certain polarization scheme.
The  $\Gamma$-modes may be divided into :
$ 4A_g +  2A_u    + B_{1g}  + 6B_{1u} + 4B_{2g} + 6B_{2u} + 3B_{3g} + 4B_{3u}$.
The schemes of the polarization patterns are given in Table I; they
result from the crystal symmetry.
$A_g$ modes show the full symmetry, their displacement parameters
hence correspond to the free parameters in the CuGeO$_3$ structure,
Ge-x, O1-x, O2-x and O2-y.
$A_u$-modes show displacements of Cu and O2 along $c$;
they are silent since shifts in neighboring chains cancel each other.
$B_{iu}$ are characterized by displacements of all atoms in the $i$-direction
with in-phase shift of equivalent atoms; in particular they contain
the starting points of the acoustic branches.
$B_{ig}$-modes are even modes which break a two-fold axis.

\clearpage

\widetext

\begin{figure} 
\includegraphics*[width=13cm,angle=90]{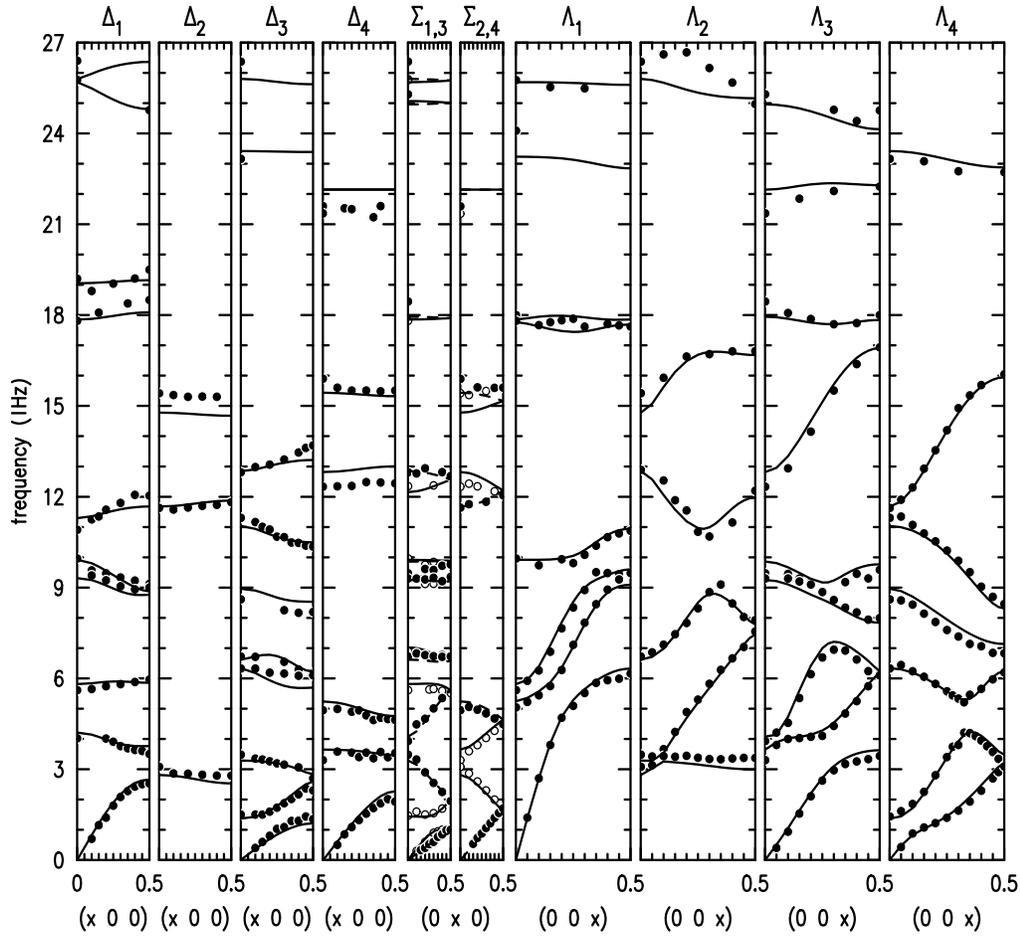}
\vskip 0.3 cm 
\caption{Phonon dispersion in \cgo :
branches are separated according to irreducible representations
along the three orthorhombic directions; circles denote
experimental points (open circles along (0x0) $\Sigma _1$- and 
$\Sigma _2$-modes) and lines the frequencies calculated by the
model.}
\end{figure}

\clearpage

\begin{table}
\begin{tabular}{| c | c |c |c |c |c  |c  |c  |c  |c  |c  |}
\hline
~~ & Cu & Cu' & Ge & Ge' & O1 & O1' & O2 & O2' & O2'' & O2'''  \\
~~ & (.5\ 0\ 0) & (.5\ .5\ 0) & (.07\ .25\ .5) & (-.07\ .75\ .5) & (.87\ .25\ 0)&(-.87\ .75\ 0) & (.28\ .08\ .5) & (-.28\ -.08\ .5) & (-.28\ .58\ .5) & (.28\ .42\ .5) \\
\hline
6\ B$_{1u}$ & A B 0 & A-B 0 &C 0 0& C 0 0& D 0 0& D 0 0& E F 0& E F 0& E-F 0& E-F 0\\
6\ B$_{2u}$ & A B 0 & -A B 0&0 C 0& 0 C 0& 0 D 0& 0 D 0& E F 0& E F 0&-E F 0&-E F 0\\
4\ B$_{3u}$ & 0 0 A & 0 0 A &0 0 B& 0 0 B& 0 0 C& 0 0 C& 0 0 D& 0 0 D& 0 0 D& 0 0 D\\
4\ A$_{g}$  & 0 0 0 & 0 0 0 &A 0 0&-A 0 0& B 0 0&-B 0 0& C D 0&-C-D 0&-C D 0& C-D 0\\
4\ B$_{2g}$ & 0 0 0 & 0 0 0 &0 A 0& 0-A 0& 0 B 0& 0-B 0& C D 0&-C-D 0& C-D 0&-C D 0\\
2\ A$_{u}$  & 0 0 A & 0 0-A &0 0 0& 0 0 0& 0 0 0& 0 0 0& 0 0 B& 0 0 B& 0 0-B& 0 0-B\\
3\ B$_{3g}$ & 0 0 0 & 0 0 0 &0 0 A& 0 0-A& 0 0 B& 0 0-B& 0 0 C& 0 0-C& 0 0-C& 0 0 C\\
1\ B$_{1g}$ & 0 0 0 & 0 0 0 &0 0 0& 0 0 0& 0 0 0& 0 0 0& 0 0 A& 0 0-A& 0 0 A& 0 0-A\\
\hline
10 $\Delta _1$  & A B 0& A-B 0& C 0 0& D 0 0& E 0 0& F 0 0& G H 0& I J 0& I-J 0& G-H 0 \\
3 $\Delta _2$   & 0 0 A& 0 0-A& 0 0 0& 0 0 0& 0 0 0& 0 0 0& 0 0 B& 0 0 C& 0 0-C& 0 0-B \\
10 $\Delta _3$  & A B 0&-A B 0& 0 C 0& 0 D 0& 0 E 0& 0 F 0& G H 0& I J 0&-I J 0&-G H 0 \\
7 $\Delta _4$   & 0 0 A& 0 0 A& 0 0 B& 0 0 C& 0 0 D& 0 0 E& 0 0 F& 0 0 G& 0 0 G& 0 0 F \\
\hline
10 $\Sigma _1$  & A B 0&-A B 0& C D 0&-C D 0& E F 0&-E F 0& G H 0& I J 0&-G H 0&-I J 0 \\
5 $\Sigma _2$   & 0 0 A& 0 0-A& 0 0 B& 0 0-B& 0 0 C& 0 0-C& 0 0 D& 0 0 E& 0 0-D& 0 0-E \\
10 $\Sigma _3$  & A B 0& A-B 0& C D 0& C-D 0& E F 0& E-F 0& G H 0& I J 0& G-H 0& I-J 0 \\
5 $\Sigma _4$   & 0 0 A& 0 0 A& 0 0 B& 0 0 B& 0 0 C& 0 0 C& 0 0 D& 0 0 E& 0 0 D& 0 0 E \\
\hline
8 $\Lambda _1$  & 0 0 A& 0 0 A& B 0 C&-B 0 C& D 0 E&-D 0 E& F G H&-F-G H&-F G H& F-G H \\
2 $\Lambda _2$  & 0 0 A& 0 0-A& 0 B 0& 0-B 0& 0 C 0& 0-C 0& D E F&-D-E F& D-E-F&-D E-F \\
9 $\Lambda _3$  & A B 0& A-B 0& C 0 D& C 0-D& E 0 F& E 0-F& G H I& G H-I& G-H-I& G-H I \\
7 $\Lambda _4$  & A B 0&-A B 0& 0 C 0& 0 C 0& 0 D 0& 0 D 0& E F G& E F-G&-E F G&-E F-G \\
\hline
17 X0X$_1$      & A B C& A-B C& D 0 E& F 0 G& H 0 I& J 0 K& L M N& O P Q& O-P Q& L-M N \\
13 X0X$_2$      & A B C&-A B-C& 0 D 0& 0 E 0& 0 F 0& 0 G 0& H I J& K L M&-K L-M&-H I-J \\
\hline
4 \tzp          & 0 0 A & 0 0 -A & 0 B 0 & 0 -B 0 & 0 0 0 & 0 0 0 & C D 0 & C D 0 & -C D 0 & -C D 0 \\
\hline
\end{tabular}
\medskip
\caption{Polarization schemes according to the crystal structure of \cgo 
~ for all $\Gamma$-modes
and for the representations along the three orthorhombic directions
and along [101] and for the \tzp -modes associated with the spin-Peierls
transition at ${\bf q}$=(0.5 0 0.5);
the first lines give the positions of 10 atoms forming
a primitive unit, the following lines show the displacements
of these atoms.
A letter at the $i$-position, signifies that this atom
is moving along the $i$-direction, a second appearance
of the same letter signifies that the second atom
moves with the same amplitude ("-" denotes a phase shift)
in the corresponding direction.}
\end{table}

\clearpage

\endwidetext

In an analogous way the crystal symmetry allows also to divide
the phonon polarization patterns for any  ${\bf q}$-value at high symmetry
points or lines in the Brillouin-zone, in particular along  the three
orthorhombic directions: $\Delta$ corresponds to [x00], $\Sigma$ corresponds
to [0x0] and $\Lambda$ corresponds to [00x].
The separation of the 30 branches according to four distinct
representations is an important simplification for the
analysis of the phonon dispersion as it may already be seen in
figure 2.

\bigskip

In the definition used here, the
$X_1$-representation always contains the longitudinal
acoustic modes with $X$=$\Delta , \Sigma$ or 
$\Lambda$.
The $X_2$-representation is characterized by the fact
that no acoustic modes correspond to it,
and $X_3$- and $X_4$-representations are given by the
transverse acoustic branches in the sequence of the polarizations,
along $a$ before along $b$ before along $c$.
The polarization patterns according to these representations are
given in Table I for the main symmetry directions and for [X0X].
The division according to the representations has significance for the
neutron scattering intensity : they yield selection rules for the
observation of the modes which allow an identification.
The relations between the irreducible representation at $\Gamma$ and
in the zone  are given by the compatibility relations shown in Table III.
For instance $A_g$-modes always correspond to the
$X_1$-representation and the $A_u$ to the $X_2$-representation.
Since CuGeO$_3$ is an insulator, polar modes exhibit Lydane-Sachs-Teller (LST) splitting,
the longitudinal frequencies correspond to starting points of $X_1$-branches
with $X$ being the polarization direction, and the transverse
frequencies to the $X_3$ and $X_4$-branches.
The symmetry further leads to a degeneration at ${\bf q}$=(0 0.5 0),
where a $\Sigma_1$-branch connects with a $\Sigma_3$-branch and
where a $\Sigma_2$-branch connects with a $\Sigma_4$-branch.

\bigskip

\begin{table}
\begin{tabular}{| r @{~:~} l @{~~~~} r @{~:~} l |}
\hline
10\ $\Delta _1 $ &  4\ A$_{g}$~+~6\ B$_{1u}$ & 10\ $\Delta _3 $ &  4\ B$_{2g}$~+~6\ B$_{2u}$ \\
3\ $\Delta _2 $ &  2\ A$_{u}$~+~ B$_{1g}$ & 7\ $\Delta _4 $ &  3\ B$_{3g}$~+~4\ B$_{3u}$ \\
\hline
10\ $\Sigma _1 $ &  4\ A$_{g}$~+~6\ B$_{2u}$ & 10\ $\Sigma _3 $ &  4\ B$_{2g}$~+~6\ B$_{1u}$ \\
5\ $\Sigma _2 $ &  2\ A$_{u}$~+~3\ B$_{3g}$ & 5\ $\Sigma _4 $ &   B$_{1g}$~+~4\ B$_{3u}$ \\
\hline
8\ $\Lambda _1 $ &  4\ A$_{g}$~+~6\ B$_{3u}$ & 9\ $\Lambda _3 $ &  3\ B$_{3g}$~+~6\ B$_{1u}$ \\
6\ $\Lambda _2 $ &  2\ A$_{u}$~+~4\ B$_{2g}$ & 7\ $\Lambda _4 $ &   B$_{1g}$~+~6\ B$_{2u}$ \\
\hline
\multicolumn{4}{|c|}{17\ X0X$_1$~:~8\ $\Lambda _1$\ +\ 9\ $\Lambda_3$ ~:~  4\ A$_{g}$~+~6\ B$_{3u}$ +   3\ B$_{3g}$~+~6\ B$_{1u}$ }\\
\multicolumn{4}{|c|}{13\ X0X$_2$~:~6\ $\Lambda _2$\ +\ 7\ $\Lambda_4$ ~:~2\ A$_{u}$~+~4\ B$_{2g}$ +   B$_{1g}$~+~6\ B$_{2u}$ } \\
\hline
\end{tabular}

\medskip
\caption{Compatibility relations in \cgo}
\end{table}

In the frame of harmonic lattice dynamics one may reduce
the equations of movement to a 3n-dimensional Eigen-value problem for each
allowed ${\bf q}$-value, see references
\cite{brueschI,brueschII},
$ \omega ^2 {\bf e} = \bar D {\bf e}$; here ${\bf e}$ is the 3n-dimensional Polarization
vector and $D$ the dynamical matrix given by :

$$ D_{\alpha, \beta}(d,d')={1 \over (m_d m_{d'})^{1/2}}\sum_{l'}\Phi_{\alpha,\beta}(0d,l'd')exp(iql') ~~~~~~(2).$$

Here the indices  $l'$ numerate the cells, $d$,$d'$ the atoms
within one cell, and $\alpha , \beta$ the three space directions.
$\Phi_{\alpha,\beta}(0d,l'd')$ are the force constants between the
atoms $d$ and $d'$ in cells shifted by $l'$ corresponding to the directions
$\alpha$ and $\beta$.
The determination of the force constants is the central problem to
analyze the phonon dispersion.

For the description of the phonon dispersion in CuGeO$_3$ we use
Coulomb-potentials with effective charges,
$V(r)\propto {Z_1Z_2e^2\over r}$ and the
repulsive forces are described by Born-Mayer potentials
$ V(r)=A\cdot exp(-r/r_o)$.
Shell charges describing a single-ion polarizability have been
introduced for all atoms with a single shell-core force constant.
The interatomic forces act on  the shells in our model.
Chaplot et al. \cite{chaplot} have reported a common model
for compounds related to the high-T$_c$ superconducting  
cuprates; these Cu-O-parameters
have been used as starting values for the description of \cgo .
The model has been initially adapted to the optical frequencies
and has been continuously refined with the inelastic neutron scattering
results.
The parameters have been fitted to the observed phonon frequencies
and to zero forces on the atoms in the equilibrium positions.
In addition,
the predictions of the model concerning the dynamic structure
factors  have been compared to the measured intensities.
All calculations were performed with the GENAX program
\cite{genax}.

We found that several features of the
phonon dispersion could only be reproduced by the inclusion of
angular forces for the Ge-O-interaction.
The angular forces, ${ d^2V \over d\alpha ^2}$,
are divided by the lengths of the two distances, in order to
yield values comparable to usual force constants.
The angular forces at the Ge-site reflect the strong
covalent character of these bonds and are expected.
However, also an angular force in the Cu-O-arrangement
gave a significant improvement, whereas,
in the HTSC cuprates no such parameters are needed.

\bigskip
\begin{table}
\begin{tabular}{c c c c | c c c }
\multicolumn{7}{c}{ionic part} \\
~ & \multicolumn{3}{c}{model I} &  \multicolumn{3}{c}{ model II} \\
\hline
ion   & Z  & Y    & K  & Z & Y & K \\
Cu & 1.77 & 4.0  & 2.8 & 1.87 & 3.91& 2.0\\
Ge & 2.26  & 0.0  & / & 1.836 & 0.0 & /  \\
O1 & -1.22 & -2.8 & 2.0 & -1.03 & -2.18 & 1.8 \\
O2 & -1.41 & -3.3 & 2.0 & -1.34 & -3.14 & 1.8 \\
~ & ~ & ~ \\
\end{tabular}
\begin{tabular}{c c c | c c }
\multicolumn{5}{c}{potentials} \\
~ & \multicolumn{2}{c}{model I} &  \multicolumn{2}{c}{ model II} \\
\hline
pair & A(eV) & r$_0$ (\AA ) & ~A(eV) & r$_0$ (\AA ) \\
Cu-O1 &  900 & 0.305 & 1008 & 0.305 \\
Cu-O2 & /    & /     & 4294 & 0.226 \\
Ge-O  & 2500 & 0.243 & /    &/ \\
O1-O1 & 1300 & 0.288 & 2000 & 0.284 \\
O1-O2 & 1800 & 0.288 & 2000 & 0.284  \\
O2-O2 & 1500 & 0.288 & 2000 & 0.284  \\
~ & ~ & ~ & ~ \\
\multicolumn{5}{c}{force constants (dyn/cm)} \\
\hline
pair  &  F     &   G    &  F & G  \\
Ge-O2 & 98846  &  30082 & 603041 & -53026 \\
Ge-O1 & 2233   &  31133 & 470090 & -37047 \\
Cu-O2 & 262013 & -29617 & ~ & ~ \\
O2-O2 (in CuO$_4$)& -15963 & 4619 & ~ & ~\\
O1-O2 & 5202   & -4591 & ~ & ~\\
O2-O2 (in GeO$_4$) & 16428 & -8400& ~ & ~ \\
~ & ~ & ~ & ~\\
\multicolumn{5}{c}{angular force constants (dyn/cm) }\\
\hline
O1-Ge-O2 & 2523  &~&8776 & ~\\
Cu-O2-Ge & 1705 &~& 4936 &~\\
Ge-O1-Ge & 10816 &~& 19754& ~\\
O2-Cu-O2 & 3753 &~& 3921 &~\\
O2-Ge-O2 & / & ~ & 10897 & ~ \\
\end{tabular}
\nobreak
\caption{Model  parameters for the description of the phonon dispersion in
\cgo ; for the explanation of the parameters see text; 
Z,Y are in electron charges; K in $10^6$dyn/cm;
in model II the uniform O-O potentials contains also
a van der Waals term -30(ev \AA$^6$)$\cdot r^{-6}$ with 
$r$ the interatomic distance in \AA.
}
\end{table}

The angular forces, however, interfere with the forces arising from the
Born-Mayer potentials. Therefore, it is impossible to describe
all identical pairs by a single potential. In a first model, model I
see Table III, additional force constants were introduced 
for nearest neighbors, the forces acting in this model
correspond hence to the sum of the direct force constants plus the
contribution from the Born-Mayer potentials.
The Cu-O2 potential as well as the van der Waals terms were completely
dropped in this model.
Model I gives a very good description of the frequency data : 
the entire set of over 700 phonon frequencies is described with a mean
deviation of 0.19\ THz.
We then attempted to simplify the model by limiting the amount of 
force constants, model II in Table III.
Only nearest neighbor Ge-O interaction is described 
by force constants, all other forces are deduced from potentials.
In addition O-O van der Waals terms and another  
angular force were introduced.
This model too gives a satisfactory description of the frequency data,
mean deviation of 0.21\ THz, but the the equilibrium forces
on the atoms are much higher.

Both \cgo -models present ionic charges 
much lower than the chemical values,
as it is typically observed.
The strong reduction, at least in case of Ge, is certainly caused
by strong covalence of the bonds.
The shell charges and forces, however, could be chosen
similar to values found in other materials \cite{chaplot}.
Due to its better description, we discuss only the results
of model I in the following.

\section{Discussion of the phonon dispersion in CuGeO$_3$ }

Several groups have performed Raman-studies on \cgo ;
the agreement between these results is good
\cite{udagawa94,popovic95b,devic94,adams88}.
In Table IV we compare the results of Udagawa et al.
\cite{udagawa94} and Popovic et al. \cite{popovic95b}
with the values obtained by inelastic neutron scattering
and the frequencies calculated with the model.
The experimental values agree within 2\% and also the agreement
with the model is satisfactory, except for the Ge-O-bond stretching
vibrations.
The latter failure is certainly due to the insufficient description of
the covalent Ge-O-bonds; within the parameter range a better
description may be obtained but at expense of the low frequency
agreement.

\begin{table}
\begin{tabular}{| c | c | c | c | c |}
\hline
~~ & Popovic et al. & neutrons & Udagawa et al.  & calculation \\
\hline
A$_{g}$  & 5.606 & 5.532& 5.516& 5.81\\
~~       & 9.953 &9.692& 9.923& 9.92\\
~~       & 17.808 & 18.13& 17.78& 17.85\\
~~       & 25.753 & /& 25.723& 25.69\\
B$_{1g}$ & 11.632 & 11.68& 11.512& 11.68\\
B$_{2g}$ & 3.478  &3.46& 3.388& 3.29\\
~~       & 6.716  &6.86& 6.626& 6.62\\
~~       & 12.801 & 12.95& 12.92& 12.85\\
~~       & 26.352 &26.71& 26.382& 25.80\\ 
B$_{3g}$ & 3.298 & 3.441& 3.298& 3.65\\
~~       & 12.322 & 12.34& 12.292& 12.82\\
~~       & 21.345 & / & 21.376 &  22.14 \\
\hline
\end{tabular}
\caption{Comparison of the Raman-results by Udagawa et al.
\cite{udagawa94} and by Popovic et al. \cite{popovic95b}
with the frequencies determined by inelastic neutron scattering
and calculations with the lattice dynamical model, all values 
are given in THz.}
\end{table}

Infrared-studies on \cgo ~ were reported by several groups
\cite{devic94,popovic95b,massa95,loosdrecht96b,li96a,damascelli97b,mcguire},
whose results correspond well for modes polarized along $b$ and $c$.
The determination of the frequencies for modes polarized along $a$
is hampered by the shape of the crystal and was attempted only by
one group \cite{popovic95b}.
Table V compares the optical results with the neutron and model
frequencies; there is only one discrepancy concerning 
an $a$-polarized mode.

\begin{table}
\begin{tabular}{| c | c | c | c |}
\hline
~~ & Popovic et al. TO/LO & neutrons & calculation \\
\hline
B$_{1u}$ & 3.927 4.017& 4.16 /& 4.081 4.212\\
~~       & 9.833 11.153& 9.3 9.6& 9.252 9.312\\
~~       & 14.33 14.66& 9.47 10.91& 9.860 11.294 \\
~~       & 18.438 19.182& 18.24 /& 17.945 19.052\\
~~       & 24.07 25.27& / / & 24.960 25.800\\
B$_{2u}$ & 1.439 1.469& 1.508 /& 1.370 1.438\\
~~       & 6.326 6.925& 6.372 /& 6.335 7.023\\
~~       & 8.544 9.234& 8.58 /& 8.961 9.419 \\
~~       & 11.302 12.412& 11.41 /& 11.03 12.156\\
~~       & 23.145 25.693& / /& 23.42 25.068\\
B$_{3u}$ &4.947 5.037& 5.109 /& 5.239 5.249 \\
~~       & 15.869 17.990& 15.73 17.27& 15.435 17.77\\
~~       & 21.586 24.074& / /& 22.15 23.23\\
A$_{u}$  & /  &3.081 & 2.811 \\
~~       & / & 15.42 & 14.77 \\
\hline
\end{tabular}
\caption{Comparison of the infrared-results by
Popovic et al.  \cite{popovic95b}
with the inelastic neutron results and calculated frequencies,
TO and LO denote the transverse and longitudinal optic frequencies,
respectively, in THz.}
\end{table}

\medskip

\begin{figure}
\begin{center}
\includegraphics*[width=0.55\columnwidth]{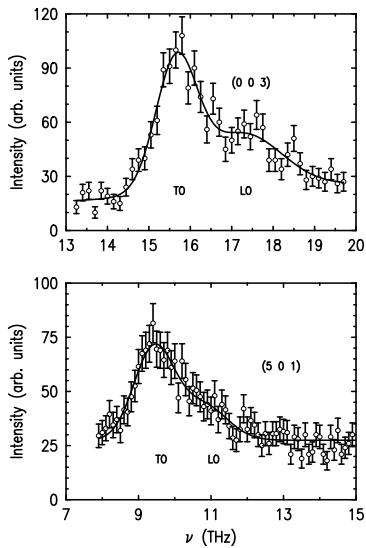}
\nobreak
\end{center}
\caption{Scans across polar modes: above  B$_{3u}$ pair,
below B$_{1u}$-pair.}
\end{figure}

Popovic et al. have interpreted a weak signal near
480cm$^{-1}$ or 14\ THz as a $B_{1u}$-mode.
This frequency is, however, not in accordance with the lattice dynamical
calculations.
The determination of the LST-splitting by inelastic neutron scattering
is in general difficult, since  the extension of the
four-dimensional resolution ellipsoid always favors
the observation of the transverse modes.
For this reason there is little chance to determine
the LO-frequency if the LST-splitting is comparable or smaller
than the resolution.
Therefore, we have attempted the measurement of the longitudinal
polar frequencies only for a few favorable cases. 
Figure 3 shows an
energy scan across the B$_{3u}$ pair at 15.7-17.3\ THz;
these frequencies confirm the infrared results.
In the lower scan one expects the $B_{1u}$-mode, assumed
by Popovic et al. \cite{popovic95b} near 14THz.
Unambiguously, this scan reveals a lower energy for this mode,
in good agreement with the model calculations.
From this scan and from the extrapolation of the branches
starting at this mode, we conclude that the TO- and LO-frequencies
of the third highest  $B_{1u}$-mode amount to 9.47 and 10.91\ THz.
The next lower  $B_{1u}$-mode is found at 9.3--9.6\ THz,
hence there is an overlap of these two pairs, which
will produce a single plateau in the reflectivity spectra at 9--11\ THz,
in addition this mode possesses little oscillator strength.
Indeed we may describe the spectrum observed by Popovic et al.
\cite{popovic95b} reasonably well on the basis of these frequencies.

The crystal symmetry only yields the schemes of the polarization patterns
as given in Table I.
For any representation with a multiplicity higher than one, one has
to know the force constants in order to determine the
polarization patterns. Figure 6 shows the polarization patterns
calculated with the lattice dynamical model.
Similar results were also reported by Popovic et al
\cite{popovic95b}; small deviations are probably due to the
misinterpretation of the  $B_{1u}$-mode.

The $A_u$-modes correspond to displacements of the Cu and O2 positions
along $c$. The mode with an opposite shift causes significant alternation
of the Cu-O2-bond distance, this mode has a high frequency of 15.4\ THz.
The in-phase vibration, where the entire ribbons
are shifted along $c$, is low in frequency at 3.1\ THz.
The distortion in the spin-Peierls phase is related to both
$A_u$-modes.

The rotation of the CuO$_2$-ribbons around the $c$-axis
corresponds to the $A_g$-mode at 9.9\ THz.
This high frequency may surprise in view of the fact that
the ribbons rotate as function of temperature around $c$
\cite{braden98a}.
The polarization pattern of the 9.9\ THz-mode lacks the
alternation of the $b$-lattice constant and the
associated shift of the GeO$_4$-tetrahedra;
without these elements the vibration is hence rather hard.

\begin{figure}[tp]
\begin{center}
\includegraphics*[width=1.05\columnwidth]{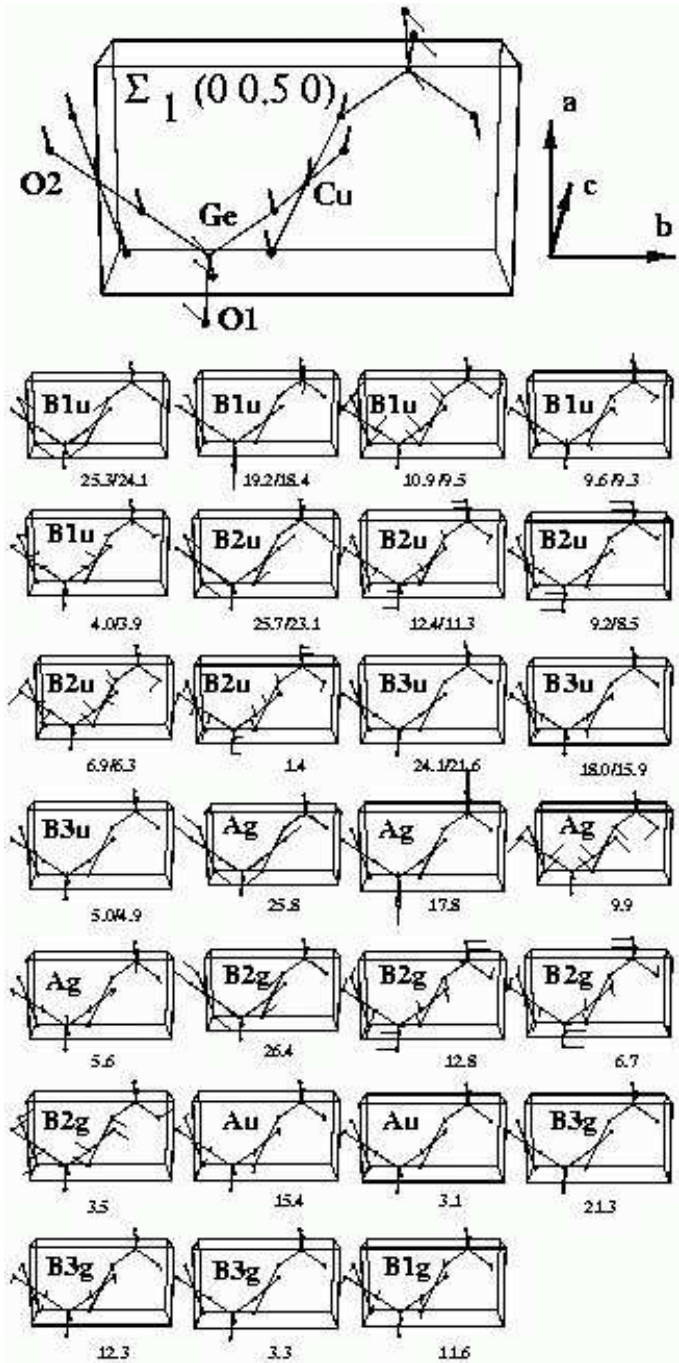} 
\smallskip
\end{center}
\caption{Polarization patterns of all $\Gamma$--modes
and their frequencies in THz.
The figure on the top illustrates the crystal structure of
CuGeO$_3$ and gives the polarization pattern of the
zone boundary mode
of the longitudinal acoustic branch (${\bf q}$=(0 0.5 0)\ ).
}
\end{figure}

Several modes correspond to rotations of the tetrahedron
chains around the $c$-axis.
The $B_{2g}$-mode at 6.7\ THz
may be described by the rotation around an axis near the middle of
the O2-O2-edges; this vibration is rather hard.
The $B_{2g}$-mode at 3.5\ THz, in contrast, corresponds to the
rotation around the O1-O1-line and is hence related to the
distortion in the spin-Peierls phase.
In this low frequency
$B_{2g}$-mode neighboring tetrahedra rotate
in the same sense, and the corresponding anti-phase rotation
is approximatively realized in the $B_{2u}$-modes at 6.3 and 1.5\ THz.
The pure anti-phase tetrahedron rotation is not an Eigen-mode
due to the stronger influence of the CuO-forces; the difference
between the two $B_{2u}$ modes concerns just the Cu-shift.

Modes with dominant Cu-contribution have lower frequencies due
to the higher mass; frequencies are particularly low for displacements
perpendicular to the Cu-O-bonds.
In the lowest  optic $B_{1u}$ mode, Cu atoms neighboring along $b$
shift almost in the direction of their connection.
The mode with the opposite phase corresponds to the lowest
$B_{2u}$-mode with a significantly lower frequency, 1.4\ THz.
This pattern allows the O2-atoms to follow the Cu-atoms
yielding an almost rigid shift of the ribbons.
Also, the tetrahedra are only slightly distorted in this pattern,
since they rotate around the $c$-axis.
In accordance with the two-dimensional character of the CuGeO$_3$
lattice \cite{braden98a} this vibration may be considered
as the transverse vibration of the zick-zack-planes.
This explains the exceptional low frequency of this mode
which is related to the low lying longitudinal acoustic
$\Sigma _1$ branch, see below.

The compared experimental and calculated phonon dispersion
shown in figure 2 indicates strong dispersion mostly
in the low frequency range.
Furthermore, the curves are steeper along $c$ than along $b$ or $a$.
This reflects the arrangement of the strong bonds, since the
phonon dispersion may be considered as the Fourier-transformation
of the force constants.
The layered structure of \cgo ~ has no covalent bonds along $a$,
therefore, the dispersion stays flat in this direction, at least
in the higher frequency range.
In consequence, one may expect that also the character of the
vibrations does not change in this direction.
In contrast, the chain configuration along $c$ will mix the
character in the Brillouin-zone along [001].

The dispersion along $b$ is determined by the coupling between
CuO$_2$-ribbons and tetrahedron chains.
The glide mirror plane in space group Pbmm transforms into the second
formula unit in the primitive cell; for vanishing coupling
one might separate the zone center modes into modes with
the two units vibrating in-  or out-of-phase but
the finite coupling results in a dispersion along $b$.
The degeneration of two branches at (0 0.5 0) corresponds in most
cases to the connection of such a pair.
The flat dispersion of many of the branches along $b$ shows
that the coupling is weak for those modes.

In the following we discuss now several particularities of the phonon
dispersion in CuGeO$_3$ which are not directly related to the
spin-Peierls transition.

{\it -- Flat longitudinal acoustic branch along $b$ and associated modes --}
Lorenzo et al. have reported a longitudinal acoustic branch in the 
$b$-direction \cite{lorenzo94}, which has been interpreted as being 
essential for the spin-Peierls transition.
The low-lying branch has been qualitatively confirmed in
later studies \cite{hirota95b,nishi95c} and also by our own results.
However, the results do not agree quantitatively.
The low-lying branch may be described within the lattice dynamical
model. The low frequencies are explained due to the lowest B$_{2u}$-mode.
From this mode a $\Sigma_1$ branch starts,
which interacts with the longitudinal acoustic branch of the same
symmetry.  The character of the acoustic vibration is hence
transferred to the optic branch, and the acoustic branch carries
the character of the B$_{2u}$-mode at 1.5THz, as it is illustrated by
the polarization pattern of the zone-boundary acoustic mode,
see top of figure 4.

The flat LA branch is hence the consequence of the
low-lying optical mode. Similar effects are also seen
in the other directions where the lowest  B$_{2u}$-mode
belongs to the  $\Delta_3$- or $\Lambda_4$-representations.
The corresponding branch interacts with the transverse
acoustic branches, $\Delta _3$ and $\Lambda_4$, respectively.

\begin{figure}
\begin{center}
\includegraphics*[height=4.1cm]{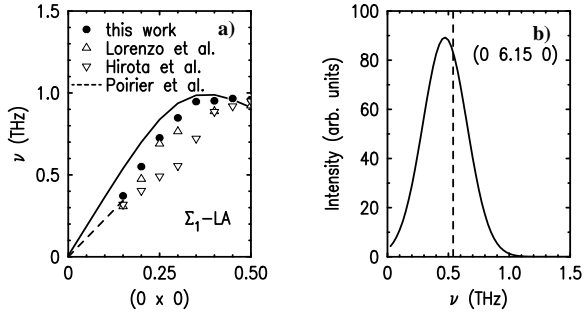}
\nobreak
\end{center}
\nobreak
\caption{a) Frequencies of the longitudinal acoustic
mode in  the $\Sigma$-direction; the measured frequencies 
of this work were
corrected for resolution effects. For comparison we also show
results of other groups \cite{lorenzo94,hirota95b}.
The dashed line corresponds to the linear extrapolation
from the elastic constant determined by
Poirier et al. \cite{poirier95c}, and the solid line to the frequencies 
calculated with our model.
b) Scan calculated by the convolution of the dispersion surface
with the spectrometer resolution for ${\bf Q}$=(0 6.15 0);
the exact value of the LA mode frequency at this ${\bf q}$-value 
is indicated by the broken line.}
\end{figure}

The flattening was interpreted  as indicating
direct relevance for the  \sppu \cite{lorenzo94}.
Damascelli et al. \cite{damascelli97a} report for the lowest B$_{2u}$-mode an
anomalous softening upon cooling, which we do not confirm.
The polarization pattern of these vibrations, however, has no
similarity with the spin-Peierls distortion in CuGeO$_3$.
The low frequencies just reflect 
a structural instability of this material.

The agreement between frequencies obtained by
different groups is poor for the LA-$\Sigma _1$-branch, see figure 5.
The convolution of the experimental resolution with the
dispersion surface may explain these differences.
Figure 5b) shows a scan calculated by the convolution of the
experimental resolution (including the mosaic spread of the sample)
with the calculated dispersion.
By several calculations of this type the resolution-induced 
correction factors have been determined and applied
to the measured frequencies.
The corrected frequencies are shown in figure 5.
Resolution effects are most likely the origin of the
bad agreement between the different neutron results.

{\it -- Soft-mode-behavior -- }
Figure 6 shows the low energy part of the phonon
dispersion of the   $\Lambda _4$- and $X0X_2$-branches.
Due to the larger amount of branches of the same symmetry,
the interpretation is rendered difficult in the [101]-direction,
but one may recognize that the branch starting at the B$_{2u}$-mode
at  6.3\ THz softens continuously through the zone,
exhibits several interactions, and ends finally at the lowest
zone-boundary frequency.
The $a$-component does not seem to be relevant for this decrease.
As shown in figure 4, the B$_{2u}$ mode corresponds to a
displacement of O2-positions in a square in the same direction
coupled with a movement of the Cu-atoms.
Along the $c$-direction the pattern changes continuously,
for ${\bf q}$=(0 0 0.5), O2-O2-edges neighboring along $c$
are moving in opposite directions, this pattern is shown in
figure 7, it may be characterized as a folding of the ribbons.
In the following this mode is called soft mode (SM).

\begin{figure}
\begin{center}
\includegraphics*[width=0.9\columnwidth]{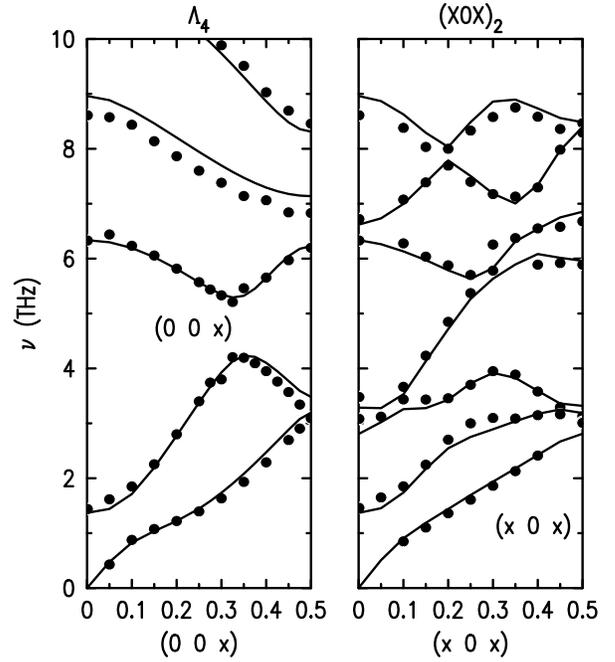}
\nobreak
\end{center}
\caption{Part of the dispersion of the
$\Lambda _4$- and $X0X_2$-branches.}

\end{figure}

The frequency decrease in these branches can be considered 
as soft-mode behavior; it points to some structural instability,
for instance this mode softens by 2.3\% upon temperature decrease
from 295 to 12\ K.
This instability in CuGeO$_3$ is still far from condensing,
but
upon substitution or pressure it might become enhanced.
The SM-instability, however, is not directly related to the
spin-Peierls transition.

The $\Lambda_4$-mode of similar frequency corresponds to a
Cu-shift, see figure 7. The mode with
opposite phase between neighboring chains is the $\Lambda_3$-mode,
in contrast to the corresponding $\Gamma$-modes,
B$_{1u}$ and B$_{2u}$, the chain coupling only weakly
affects the zone-boundary modes.

{\it -- Behavior of the highest Ge-O-branches --}
The dispersion of the highest Ge-O-branches is not fully understood.
In general the description of these branches by our model is poor
due to the insufficient description of the covalent forces, see above.
The behavior of the highest
$\Lambda_2$- and $X0X_2$-branches, shown in figures 2 and 11,
is surprising. Again, the influence of the $a$-component is negligible.
The bending down of the branch towards the zone-boundary
is reproduced by the model only in general.
We do not think that this behavior results from magneto-elastic coupling,
but it appears more likely to be an intrinsic property of the
tetrahedron chains. "ab-initio" calculations should give
more insight into this problem.

\begin{figure}[tp]
\begin{center}
\includegraphics*[width=5cm,angle=-90]{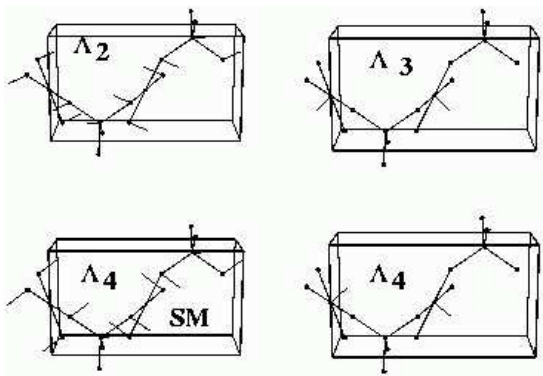}
\smallskip
\end{center}
\smallskip
\caption{Polarization-pattern of the four lowest modes
at ${\bf q}$=(0 0 0.5).}
\end{figure}

{\it -- Further predictions of the lattice dynamical model --}
Table VI compares the elastic constants determined by ultra sound
and by Brillouin scattering techniques 
with those calculated by the model.
The first ultrasound measurements were limited to the diagonal terms
\cite{poirier95c,poirier2,saint-paul} and a complete set of 
elastic constants was determined only by Ecollivet et al. using 
Brillouin scattering \cite{ecollivet}.
There is some scattering in the experimental data obtained with the 
ultra-sound techniques. The model values compare best to the combined 
study in reference \cite{ecollivet}.
The lattice model predicts 
C$_{22}$ to be significantly higher than the experimental value, 
as may be also seen in figure 6.
The model seems to slightly
underestimate the interaction between the optic
and acoustic branches near the zone center.
Upon increase of the weight of the LA modes one may
obtain better agreement for the elastic constants but at the  expense
of the description at higher energies, which has been considered to be more
relevant.

\bigskip

\begin{table}
\begin{tabular}{ c c c c c c c c c c}
 ~ & c11 & c22&c33&c44&c55&c66&c12&c13&c23 \\
US-a & 0.624 & 0.372 & 3.264 &/&/&/&/&/&/ \\
US-b & 0.74 & 0.21 & 3.32 &/&/&/&/&/&/ \\
US-c & 0.66 & 0.345 & 2.79 &/&/&/&/&/&/ \\
US-d & 0.71    & 0.345 & 3.43 &   0.37  & 0.33   & 0.22     &/&/&/ \\
BR    & 0.64   &0.376& 3.173 &    0.353 & 0.353 & 0.184    & 0.321 & 0.469 & 0.227 \\
model &  0.823 &0.500& 3.457 &    0.408 & 0.435 & 0.165    & 0.297 & 0.403 & 0.223 \\
\end{tabular}
\nobreak
\caption{Comparison of the experimentally determined elastic constants
(ultra-sound : US-a Poirier et al. \cite{poirier95c},
US-b Poirier et al. \cite{poirier2},
US-c Saint-Paul et al., \cite{saint-paul},
US-d Ecollivet et al., \cite{ecollivet};
Brillouin scattering : BR Ecollivet et al., \cite{ecollivet}) 
with those calculated with the lattice dynamical model;
elastic constants are given in $10^{12}dyn/{cm^2}$
.}
\end{table}

\begin{figure}[tp]
\begin{center}
\includegraphics*[width=0.8\columnwidth]{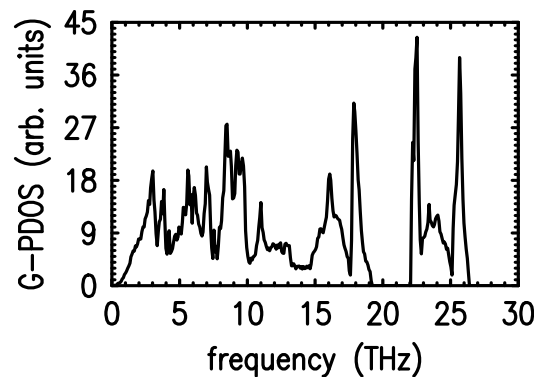}
\nobreak
\end{center}
\smallskip
\nobreak
\caption{GPDOS calculated with the lattice dynamical model.}
\end{figure}

The phonon density of states, PDOS,  has been studied by Arai et al.
\cite{arai94} and by Fujita et al. \cite{fujita96a}
by inelastic neutron scattering on polycrystalline samples.
In figure 8 we show the PDOS weighted with the atomic cross section
divided by the mass, the generalized PDOS (GPDOS),
calculated with the lattice dynamical model;
the agreement with the experiment is at most qualitative.
However, the good agreement of the model with the optic frequencies and
with the dispersion curves determined here excludes the experimental
GPDOS. We think that the deviation results from insufficient
correction of multi-phonon processes.
Also the strong temperature dependences point to such effects
\cite{arai94,fujita96a}.
Neither our studies, nor the optical experiments nor the
similar work by Nishi et al. \cite{nishi95c}
give evidence for a strong temperature dependence of the phonon
spectrum in CuGeO$_3$.
In addition, the measured GPDOS presents huge intensity above
the cut-off value of 26\ THz, which however, is already fixed
by the optics studies.

From the calculated PDOS one may obtain the specific heat at constant
volume. Most of the published work on the
specific heat deals with the anomaly at the spin-Peierls transition
\cite{kuroe94b,sahling94,liu95b,weiden95}.
Only Weiden et al. \cite{weiden95} report a measurement up to room temperature
which is compared in figure 9 with the results of the calculations.
The difference between specific heat at constant volume and at constant
pressure has been corrected empirically,
indicated by the solid and dashed lines in figure 9.
Since the coefficient of thermal expansion enters the difference
in second order, the correction is extremely small, it remains
almost in the width of the lines.
Surprisingly, the measured specific heat is always about 12\% higher than
the calculated one; at least at higher temperature this can not be attributed
to magnetic contributions.
Again, it seems unlikely that the lattice dynamical model is that wrong;
for instance the large number of high optic frequencies
unambiguously indicate that the specific heat at room temperature
is far from its saturation value corresponding to the
Dulong-Petit law. 
A larger specific heat might arise from strong anharmonicity,
as in the classic example of AgI where a sub-lattice melts  \cite{brueschI}. 
The anharmonicities in CuGeO$_3$ appear to be, however, too small
for such an explanation.

\begin{figure}
\begin{center}
\includegraphics*[height=4cm]{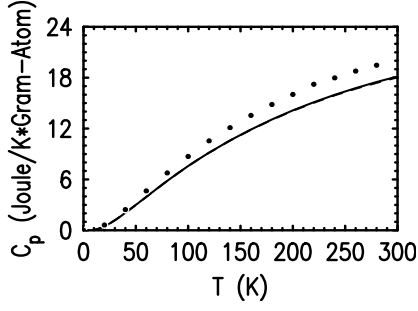}
\nobreak
\end{center}
\smallskip
\nobreak
\caption{Comparison of the experimental specific heat (points, taken from
reference
\cite{weiden95}) with the calculated ones; the solid line corresponds
to $c_p$and the dashed one to $c_V$.}
\end{figure}

Another check of the model may be obtained through the comparison with the
experimental Debye-Waller factors, see Table VII.
The agreement at room temperature is very good, and also the
low temperature coefficients, which are more difficult to determine,
are well described \cite{braden98a}.

\begin{table}
\begin{tabular}{c c l | r c c}
atom &  exp. & cal. & atom & exp. & cal. \\
\hline
Ge-U11 &  70(3)& 55  &  O1-U11& 81(4)& 64\\
~~~-U22& 101(3)& 90  & ~~~-U22&129(4)& 117 \\
~~~-U33&  40(3)& 31  & ~~~-U33& 48(3)& 42\\
Cu-U11 & 115(3)& 100 &  O2-U11& 144(3)& 127\\
~~~-U22& 142(2)& 116 & ~~~-U22& 178(3)& 161\\
~~~-U33&  46(3)& 42  & ~~~-U33& 61(3)& 56\\
~~~-U12&  50(2)& 37 & ~~~-U12& 79(2)& 69\\
\end{tabular}
\smallskip
\caption{
Debye-Waller-coefficients in \cgo ~ at room temperature :
comparison of experimental (average of x-ray and neutron
results from \cite{braden96})
and calculated values, in 10$^{-4}$\AA $^2$.}
\end{table}

\begin{figure}[t]
\begin{center}
\includegraphics*[width=5cm,angle=-90]{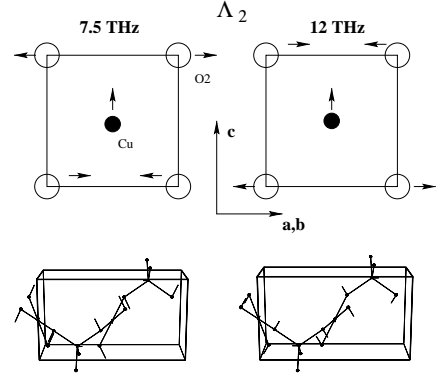}
\smallskip
\end{center}
\smallskip
\caption{Polarization patterns of the two modes with
Cu-shift along $c$ at ${\bf q}$=(0 0 0.5); the upper two 
pictures show a projection of a single CuO$_4$-square.}
\end{figure}

\section{modes in relation to the spin-Peierls transition }

The structural part of the spin-Peierls transition is characterized by a
symmetry reduction in the crystal structure from Pbmm to Bbcm.
The relevant phonon modes may be identified with the help
of group theory. In the notation
by Stokes and Hatch \cite{stokes88}
these modes are labeled \tzp ~ at ${\bf q}$=(0.5 0 0.5),
The\tzp -modes  are characterized by four independent parameters 
in their polarization pattern
which reflect the four additional structural parameters in the
dimerized structure.
The corresponding irreducible representation is one-dimensional and has a
multiplicity of four.
The \tzp -modes have been discussed in detail in
references \cite{braden98b,werner98} including their temperature dependence.
In this section we want to give some further information
on the dispersion of the branches connected to
these four modes directly involved.
The lowest \tzp -mode at 3.12\ THz is B$_{2g}$-like associated with a 
rotation of the O2-O2-edges around the $c$-axis; it modulates the
O2-O2-Ge angle, $\delta$, and, therefore, also the magnetic interaction J.
The second lowest mode at 6.53\ THz is associated with the modulation 
of the Cu-O2-Cu bond angle $\eta$, which determines strongest the 
magnetic interaction. This $\eta$-mode seems to be the one most important
for the spin-Peierls transition.
The higher frequency \tzp -modes at 11 and 25\ THz correspond to 
the Cu-O2- and Ge-O bond stretching and are less relevant for the
spin-Peierls transition.
The \tzp -representation corresponds to the propagation vector of
${\bf q}$=(0.5 0 0.5), therefore, it is interesting to analyze
the dispersion in the [101]-direction, see figure 11.

\begin{figure}[tp]
\begin{center}
\includegraphics*[width=0.9\columnwidth]{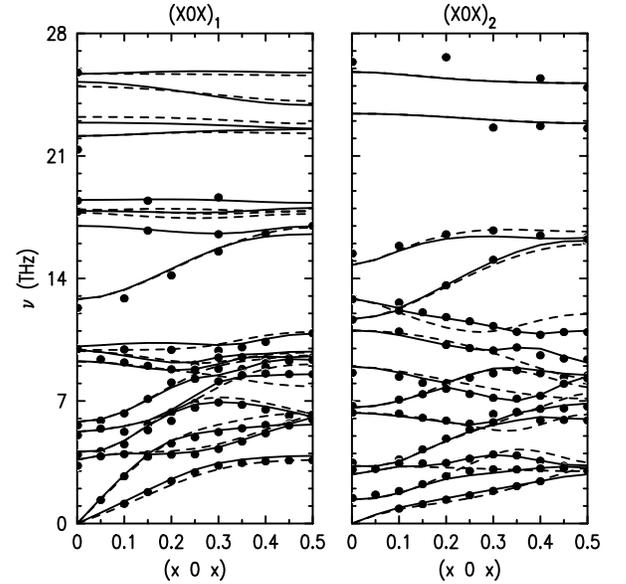}
\nobreak
\end{center}
\smallskip
\nobreak
\caption{Dispersion of the branches in the [101]-direction. The
symbols denote measured frequencies and the solid lines those calculated
by the lattice dynamical model. 
Dashed lines denote the dispersion of corresponding
$\Lambda$-branches neglecting the influence of the        
$a^*$-component.}
\end{figure}

{\it -- [101]-direction and identification of the relevant modes --}
The $\Gamma$-modes whose polarization patterns are
closest to the structural distortion in the
spin-Peierls phase are the A$_u$ and B$_{2g}$-modes where
branches of $\Lambda_2$-symmetry start in the
$\Lambda$-direction.
As may be seen in the polarization schemes given in Table I, the
$\Lambda_2$-representation allows the $z$-shift of Cu and
the $x,y$-shifts for O2.
These Cu and O-characters of these modes 
are well separated near the
zone center but they mix in the zone.
Along the [101]-direction the modes may be divided only according to two
different representations, 17 $X0X_1$- and 13 $X0X_2$-modes.
The large number of modes of the same symmetry
renders the interpretation of the data rather difficult.
The compatibility relations show that the 
\tzp -modes correspond to the zone-boundary modes
of the $X0X_2$-branches, see Table I.

At the zone-boundary (0 0 0.5) there are two vibrations
where the Cu  moves along $c$, the upper one is isolated
in frequency near 12\ THz and the lower one near 7.7\ THz.
As may be seen in figure 10, the difference in these modes consists
in the phases of Cu and O2-displacements.
In the higher mode, Cu moves towards the shorter O2-O2-edge yielding
a strong alternation of the Cu-O-bond lengths and, therefore, the
higher energy. In contrast, for the lower mode Cu shifts towards the
longer edge yielding an alternation of the bond angle.
Besides the $a$-component this movement represents the main
part of the structural distortion in the dimerized phase.
These modes are called $\eta$-modes in the following.

The B$_{2g}$-mode near 3.5\ THz corresponds to the rotation of the
edges of the CuO$_4$-squares and hence to the second dominant
feature of the spin-Peierls distortion.
A flat $\Lambda_2$-branch is starting at this mode, which ends at the
zone-boundary at a mode with opposite twisting of edges neighboring along $c$.
Again, only the $a$ component is missing for the description of the
distortion in the dimerized phase. In general one may arrive at the
relevant \tzp -modes by either analyzing the path from (0 0 0.5) to (0.5 0 0.5)
where there is only little dispersion or studying the [101]-dispersion,
see below and reference \cite{braden98b}.

Figure 11 shows the phonon dispersion  in the [101]-direction.
Again the dispersion is well described by the model.
From the compatibility relation in Table II one recognizes that
$X0X_1$-branches correspond to the combination of
$\Lambda _1$-  and $\Lambda _3$-branches in the [001]-direction,
and $X0X_2$-branches correspond to
$\Lambda _2$- and $\Lambda _4$-branches.
In figure 11 we show the corresponding $\Lambda$-branches as
broken lines neglecting the $x$-component of the wave-vector
(i.e. in a projection).
The resemblance shows that the $a^*$-component influences the
frequencies only slightly, in agreement with the flat dispersion along
$a$ and the lack of any covalent bond in this direction.
However, the influence of the $a^*$-component
may not at all be neglected for the interpretation of the intensities.
Along the  $\Lambda$-direction one observes only  $\Lambda_1$-branches
at ${\bf Q}$=(0 0 2+x); the intensities of $\Lambda_3$-modes
are suppressed, similar to an elastic extinction rule.
In the  [101]-direction, however, one may also observe the $X0X_1$-modes
corresponding to the  $\Lambda _3$-modes at ${\bf Q}$=(x 0 2+x)
with comparable intensities.
The observation that structure factors are much more sensitive than
the intensities corresponds to the experience with small
structural distortions \cite{reich-bra-mang}.

Due to the larger number of modes of same symmetry the mixing 
of the characters is even more important in the [101]-direction.
Nevertheless, it has been possible to clearly identify the \tzp -modes
at ${bf q}$=(0.5 0 0.5) in reference \cite{braden98b}. 
The \tzp -mode rotating the O2-edges is found at 3.12\ THz,
the most relevant $\eta$-modulating mode at 6.53\ THz, and the
two highest \tzp -modes exhibit frequencies of 11.2 and 24.9\ Thz,
respectively, see reference \cite{braden98b}.

{\it -- Comparison with Raman-studies in the dimerized phase--}
Raman-studies below \tsp ~ have been very early performed 
by Udagawa et al.
\cite{udagawa94}
and by Sugai et al. \cite{sugai93}; they show
additional peaks in the spin-Peierls phase.
In the meanwhile there have been many publications on 
these intensities, which agree concerning the 
experimental findings but not on the interpretation
\cite{sugai93,udagawa94,kuroe94a,ogita96a,kuroe96a,gros97a,
loa96,loosdrecht96a,lemmens96a,ogita96b,muthukumar97a}.
All groups find five additional peaks; additional much weaker
intensities are reported in reference \cite{loa96,ogita96b} 
and appear to arise from some different mechanism.

The lowest intensity near  30cm$^{-1}$ is of magnetic origin.
All Raman-interpretations agree that the highest two 
strong peaks correspond to folded phonons, they perfectly agree
to our results on the two highest \tzp -modes.
But also the two peaks at lower frequencies
which were interpreted controversely may be 
identified as \tzp -modes, since their frequencies 
perfectly agree with those of the 
B$_{2g}$-like and $\eta$- \tzp -modes.
The neutron scattering studies reveal a strong broadening 
of the  $\eta$-mode in good agreement with the Raman-experiments;
Lemmens et al. \cite{lemmens96a} report a width of 16cm$^{-1}$.
In addition, it has been found that just the  Raman-intensity corresponding 
to the $\eta$-mode can be observed a few degrees K above the \sppu .
\cite{lemmens96a,ogita96b}. This underlines once more
the exceptional role of the $\eta$-mode in the 
spin-Peierls transition.

{\it -- Dispersion of the branches connected to the \tzp -modes --}
In general the dispersion is rather flat along the $a$-direction.
This is also valid for branches starting at the \tzp -modes
and passing along the path (0.5-x 0 0.5) on the zone boundary.
The endpoint corresponds to the zone-boundary of the $\Lambda$-direction.
The small influence of the $a^*$component is already illustrated in 
figure 11 and was used to discuss the character of the \tzp -modes.
The modes at ${\bf q}$=(0 0 0.5) may be further divided according to 
symmetry . In particular,
there is a representation of multiplicity four which corresponds to the
\tzp -modes at (0.5 0 0.5).
The frequencies of these four modes lie at
3.29, 7.69, 12.1 and 24.9\ THz, compared to the \tzp -frequencies
at 3.12, 6.53, 11.2 and 24.9\ THz at room temperature.

The (0 0 0.5)-mode corresponding to the B$_{2g}$-like 
\tzp -mode turning the O2-edges
is only slightly harder than the \tzp -mode, at room temperature. 
At low temperature,
the difference is even smaller since the (0 0 0.5)-mode
hardens slightly to 3.34\ THz, whereas the \tzp -mode
is significantly shifted due to the magneto-elastic coupling
\cite{braden98b}.

The difference in frequency  for the modes modulating the
Cu-O-Cu-bond angle, $\eta$, the main feature of the
spin-Peierls distortion in CuGeO$_3$, is more pronounced, about 1\ THz.
This dispersion explains the doubling of the lattice in the
$a$-direction; the distortion with modulation along $a$ just
requires less structural energy.
For these $\eta$-modes too, the stronger frequency increase 
upon cooling is observed
for ${\bf q}$=(0.5 0 0.5); but also the (0 0 0.5)-mode frequency increases 
significantly, by 1.5\% .

The second highest \tzp -mode exhibits a significant frequency increase
towards (0 0 0.5), whereas the Ge-O-modes are little influenced by the
$a^*$-component.

The dispersion along the $b$-direction on the zone-boundary (0.5 x 0.5)
is comparable to that in the $\Sigma$-direction.
In particular, there is a degeneration of two modes at (0.5 0.5 0.5).
Again, two (0.5 0 0.5)-modes connect which correspond to the phase
shift between the two formula units.
Figure 12 shows the dispersion along this path, and figure 13
indicates the polarization patterns of several (0.5 0 0.5)-modes
relevant for the following discussion.

\begin{figure}[tp]
\begin{center}
\includegraphics*[height=7cm]{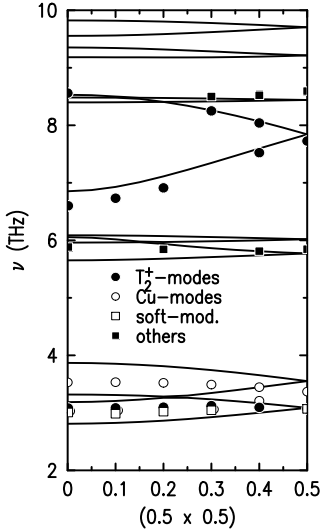}
\nobreak
\end{center}
\smallskip
\nobreak
\caption{Dispersion of branches connected to the
\tzp -modes along the path (0.5 $x$ 0.5) at room temperature ;
symbols denote experimental frequencies and lines calculations.}
\end{figure}

There are four modes in the range  3--4\ THz at ${\bf q}$=(0.5 0 0.5): 
the \tzp -mode, two
Cu-modes, associated to the lowest optic B$_{1u}$- and B$_{2u}$-frequencies
and the soft mode.
As in the $\Sigma$-direction, the two Cu-modes connect.
For the \tzp -mode the situation is more complicated :
considering the scheme of a single CuO$_2$-ribbon, one might
expect an opposite-phase mode, where along $b$ neighboring O2-O2-edges
rotate in the same sense concerning their angle to the $a$-axis.
This movement, however, would result in a strong modulation of
the Ge-O-bonds causing much higher frequency.
On the contrary one may consider this \tzp -mode as a rotation
of the GeO$_4$-tetrahedra around the O1-O1-line.
In the \tzp -mode two tetrahedra neighboring along $b$
turn both clockwise or both anti-clockwise;
the movement with different phase -- one clockwise the other
anti-clockwise -- results in a completely different
displacement scheme concerning the CuO$_2$-ribbons,
since both O2-positions in an O2-O2-edge perpendicular to
$c$ move then in the same direction.
This pattern corresponds to the soft mode at (0.5 0 0.5), see figure 13.
The lowest \tzp -mode connects hence with the soft mode along the
$b$-direction but looses the modulation of the magnetic interaction
along this path.
The anomalous frequency increase upon cooling \cite{braden98b}
observed for the \tzp -mode
has been followed along  $b^*$; the frequency hardening continuously
decreases till (0.5 0.5 0.5);
in contrast the soft-mode even softens by 2.3(3) \%
upon cooling from room temperature to 12\ K.

For the $\eta$-mode
the Ge-O-forces do not play an important role;
therefore, one may easily follow these modes along  $b^*$.
One finds a significant frequency enhancement; 
the corresponding mode is found
at 8.56\ THz, its polarization pattern, see figure 13, exhibits
a smaller Cu-displacement than that of the \tzp -$\eta$-mode.
In the range 5-6\ THz, there are several flat branches
included in figure 12 which are not related to the
spin-Peierls distortion.
The temperature dependence of the bond angle modulating
modes close to (0.5 0 0.5) could not be studied, since in this
region additional magnetic intensity appears upon cooling
\cite{braden-prl2}.
For $q_b$=0.5 no temperature dependence has been detected,
also the intensities near 5.8\ THz are almost 
temperature independent.

\begin{figure}[tp]
\begin{center}
\includegraphics*[width=5cm,angle=-90]{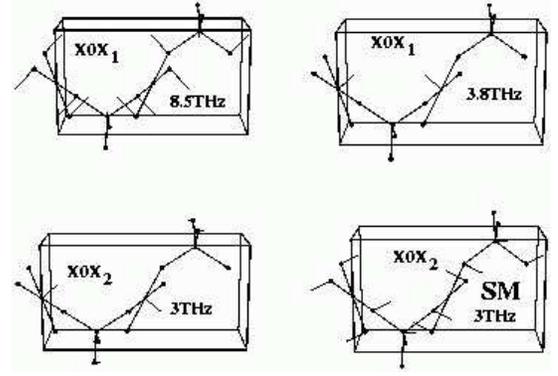}
\smallskip
\end{center}
\caption{Polarization patterns of four modes at ${\bf q}$=(0.5 0 0.5),
which are related to the \tzp -modes at 6.6\ THz ($\eta$-mode) and at
3.1\ THz (B$_{2g}$-like mode).}
\end{figure}

The dispersion of the branches connected to the two lower
\tzp -modes explains the propagation vector of the
structural distortion in the spin-Peierls phase, ${\bf q}$=(0.5 0 0.5).
The $b$-component is fixed through the Ge-O-forces, which
shift the mode rotating the O2-O2-edges to very high energies
and which also case a sizeable slope of the $\eta$-modulating
branch.
The weaker inter-plane forces lead to the doubling
of the $a$-parameter mainly due to their influence on the
$\eta$-modes.
The structural distortion involved in the spin-Peierls transition
has, therefore, a clear three-dimensional character.
Therefore, a RPA treatment of the spin-phonon coupling in the
transition appears to be justified \cite{werner98}.

The dispersion of the phonon branches should reflect
the extension of the critical scattering along the three
orthorhombic directions.
In a classical soft-mode transition one may
expand the frequency of the involved mode
at the propagation vector of the structural distortion,
${\bf q_0}$, here (0.5 0 0.5) with
${\bf q}={\bf q_0}+{\bf q_a}+{\bf q_b}+{\bf q_c}$ :

$$ \omega ({\bf q})=\omega _0 + c_a \cdot {\bf q_a}^2 + c_b \cdot {\bf q_b}^2 +
      c_c \cdot 
{\bf q_c}^2~~~~(4). $$

Linear terms  do not exist, since there must be
a frequency minimum at ${\bf q_0}$.
The ratio of the constants  $c_a,c_b,c_c$
should correspond to the ratio of the correlation lengths
of critical scattering
in the three directions.
In \cgo ~ this analysis is not obvious since there is no
softening and since at least the correlation length
along $c$ is dominated by the magnetic interaction.
Furthermore, the fitting of the experimental dispersion by a quadratic
expansion is not very satisfactory; one obtains
$c_a$= 2.6\ THz/\AA$^2$, $c_b$=8.0\ THz/\AA$^2$ and  $c_a/c_b$= 0.3,
for the $\eta$-mode which is the most relevant one.

The observed correlation lengths decrease very rapidly
above the transition; only a few degrees K above it they amount to
a few lattice constants; therefore, rather large  ${\bf q_a}$ and
${\bf q_b}$-values are involved, where a quadratic expansion
is certainly no longer valid.
Nevertheless, the ratio  between the correlation lengths
determined by Schoeffel et al. \cite{schoeffel96a}
${\xi _a \over \xi _b}\sim {1/4}$ may be qualitatively explained
by the dispersion of the branches around the \tzp -$\eta$-mode.

\section{Possibility of independent order parameters}

\begin{figure}[tp]
\begin{center}
\includegraphics*[width=0.6\columnwidth]{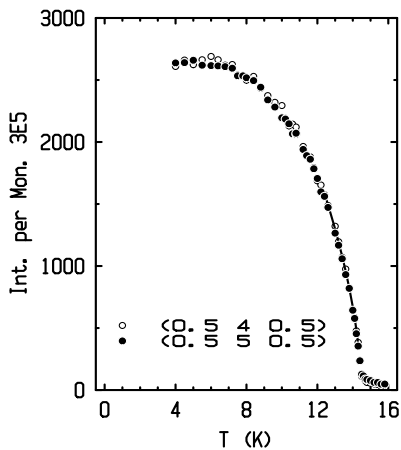}
\nobreak
\end{center}
\smallskip
\nobreak
\caption{Intensity of the two superstructure reflections at 
${\bf Q}$=(0.5 3 0.5)  and (0.5  4 0.5) as function of temperature.
The two intensities reflect approximately the distortion of the
two lowest \tzp -modes. The line corresponds to a fit by power law.}
\end{figure}

Since the structural distortion in the spin-Peierls phase does not
correspond to a single phonon Eigen-mode, one might speculate whether
the low temperature phase is described by a single order parameter
or whether more than one parameters are needed in order to 
describe the temperature dependence of the distortion.
We have analyzed this problem by measuring two superstructure 
reflections which are not sensitive to the same features of the
distortion. Intensity at (0.5 3 0.5) is mainly related to the 
distortion corresponding to the lowest \tzp -mode with 
B$_{2g}$-character, and intensity at (0.5 4 0.5) reflects the 
$\eta$-modulation. In spite of their different sensitivities
these two super-structure exhibit exactly the same temperature
dependence as it is shown in figure 14.
Fitting a critical behavior, $ I \propto (T_{SP}-T)^{2\beta}$, 
to the temperature dependence 
of these superstructure intensities yields
a critical exponent of $\beta$=0.30(1) in good agreement with 
previous results \cite{harris94,harris95,lumsden96b}
and in particular with the thermal expansion study
\cite{winkelmann95}.
The scaling between the two superstructure reflections remains
also valid slightly above the spin-Peierls transition, indicating
that the whole structural transition may be described by a single
order-parameter.

\section{Conclusion}

The various inelastic neutron scattering studies have yielded a
detailed understanding of the lattice dynamics in CuGeO$_3$.
The developed model not only describes the observed 
dispersion curves, but it allows the interpretation of 
Raman and infrared-frequencies, elastic constants and 
anisotropic Debye-Waller parameters. 
It further predicts the PDOS and the phononic part of the 
specific heat, on which only little experimental information
exists till today.

Only the extensive study of the phonon dispersion has allowed the
identification of those modes directly involved in the transition
\cite{braden98b}.
The dispersion of the branches connecting with these modes 
clearly explains the occurrence of the structural transition 
at ${\bf q}$=(0.5 0 0.5), since the phonon frequencies of modes
modulating the bond-angle $\eta$ are lowest at this ${\bf q}$-value.
The pronounced dispersion of the phonon branches involved in the
spin-Peierls transition clearly illustrate a three-dimensional
character of the structural part of the transition.
The influence of the magneto-elastic coupling on the 
temperature dependence of phonon frequencies should 
occur not only in CuGeO$_3$ but also in 
materials with related structure, like for example the
spin-ladder compounds.

CuGeO$_3$ is amongst the most complex materials where the phonon
dispersion has been analyzed in that detail. It appears therefore to 
be a promising candidate for the extension of current ``ab initio''
techniques to complex materials.

{\bf Acknowledgments.}  Work at Cologne University was supported
by the Deutsche Forschungsgemeinschaft
through the Sonderforschungsbereich 608.

* electronic mail : braden@ph2.uni-koeln.de

\end{document}